\documentclass{kluwer}    
\usepackage{graphicx}
\newdisplay{guess}{Conjecture}

\begin{document}                                                                                   
\begin{article}
\begin{opening}         
\title{Of Variability, or its Absence, in HgMn Stars}
\author{Sylvain \surname{Turcotte}}  
\runningauthor{Turcotte \& Richard}
\runningtitle{Variability in HgMn Stars}
\institute{Lawrence Livermore National Laboratory, L-413, P.O. Box 808, Livermore, CA 94551,
USA}
\author{Olivier \surname{Richard}}  
\institute{D\'epartement de physique, Universit\'e de Montr\'eal, Succ.
Centre-Ville, Montr\'eal, Qu\'ebec, Canada}
\date{August 31, 2002}

\begin{abstract}
Current models and observations of variability in HgMn stars disagree.
We present here the models that argue for 
pulsating HgMn stars with properties similar to those of Slowly
Pulsating B Stars. The lack of observed variable HgMn stars suggests
that some physical process is missing from the models. Some
possibilities are discussed.
\end{abstract}

\end{opening}           

\section{Introduction}  

HgMn stars are late B type stars featuring large and varied abundance
anomalies. They are slowly rotating, non-magnetic and mostly young stars
and are thought to be the purest examples of stars undergoing
microscopic diffusion without the complications posed by mixing
processes in cooler stars with convective upper layers
\cite{VauclairVauclair82}. 
From a pulsation point of view they are as of yet
entirely constant with a maximum photometric variability estimated at 
less than 5 mmag \cite{Adelman98}. There has only been a few studies of spectral
variability in HgMn stars leading to suggestions of some variability in
Hg lines \cite{Adelmanetal02}. There are no serious suggestions of pulsations in line profile
variations at this time (see e.~g. Turcotte et~al. in these proceedings).
There have been claims of rotational variability linked to
possible magnetic fields or abundance spots
\cite{Adelmanetal02}.
\begin{figure} 
\centerline{\includegraphics[width=15pc]{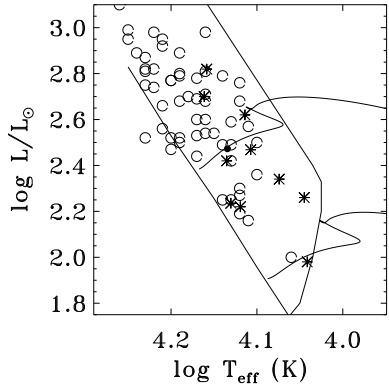}}
\caption[]{HR Diagram showing some HgMn stars (asterisks), some SPB stars (open circles), and the
           theoretical limit of the SPB instability region \cite{Pamyatnykh99}. The
           evolutionary tracks for models of 3 and 4~M$_\odot$ are shown and the 
           location of the model which will be discussed here is
           indicated by the filled circle.}
\label{fig:HRD}
\end{figure}

HgMn stars share a part of the HR diagram in which pulsating stars are
present as shown in Fig.~\ref{fig:HRD}. Those stars, the Slowly Pulsating B stars (SPBs), are in many
ways similar to HgMn stars, many seem to be slowly rotating, but differ in 
that they are chemically normal. In addition, there is a theoretical
expectation that pulsations should be driven in HgMn stars at least as
much as those in SPBs, suggesting that detailed studies of the stability
of HgMn stars, both observationally and theoretically, is likely to
yield new information on the structure and dynamics of B stars.

\section{Theoretical Expectations}

Sophisticated models of diffusion in stars can now be made for cool B
stars following the work of \cite{Richeretal00} and
\cite{Richardetal01}.
In those papers, it has been demonstrated that diffusion in A
and B stars can lead to a substantial increase in opacity in the region
where heavy elements contribute the most to it, at a temperature 
around 200\,000~K. In some cases, when metals are allowed to accumulate
enough, a convection zone can form in that region.
It is also the same region in which pulsations in B stars 
(both SPBs and hotter $\beta$~Cephei stars) are 
driven \cite{Pamyatnykh99}.

\begin{figure}
 \tabcapfont
 \centerline{%
 \begin{tabular}{c@{\hspace{2pc}}l}
  \includegraphics[width=2in]{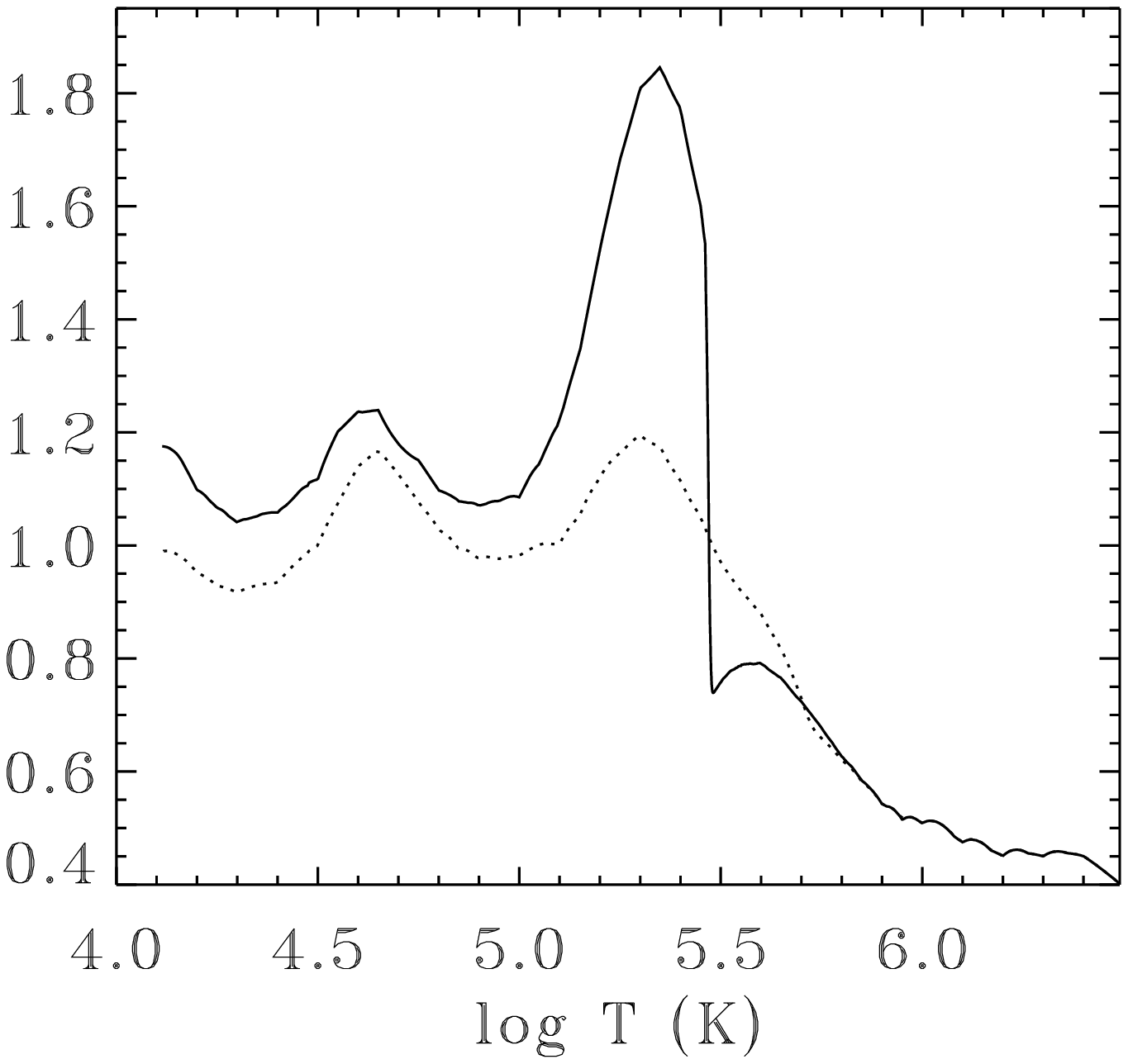} &
  \includegraphics[width=2in]{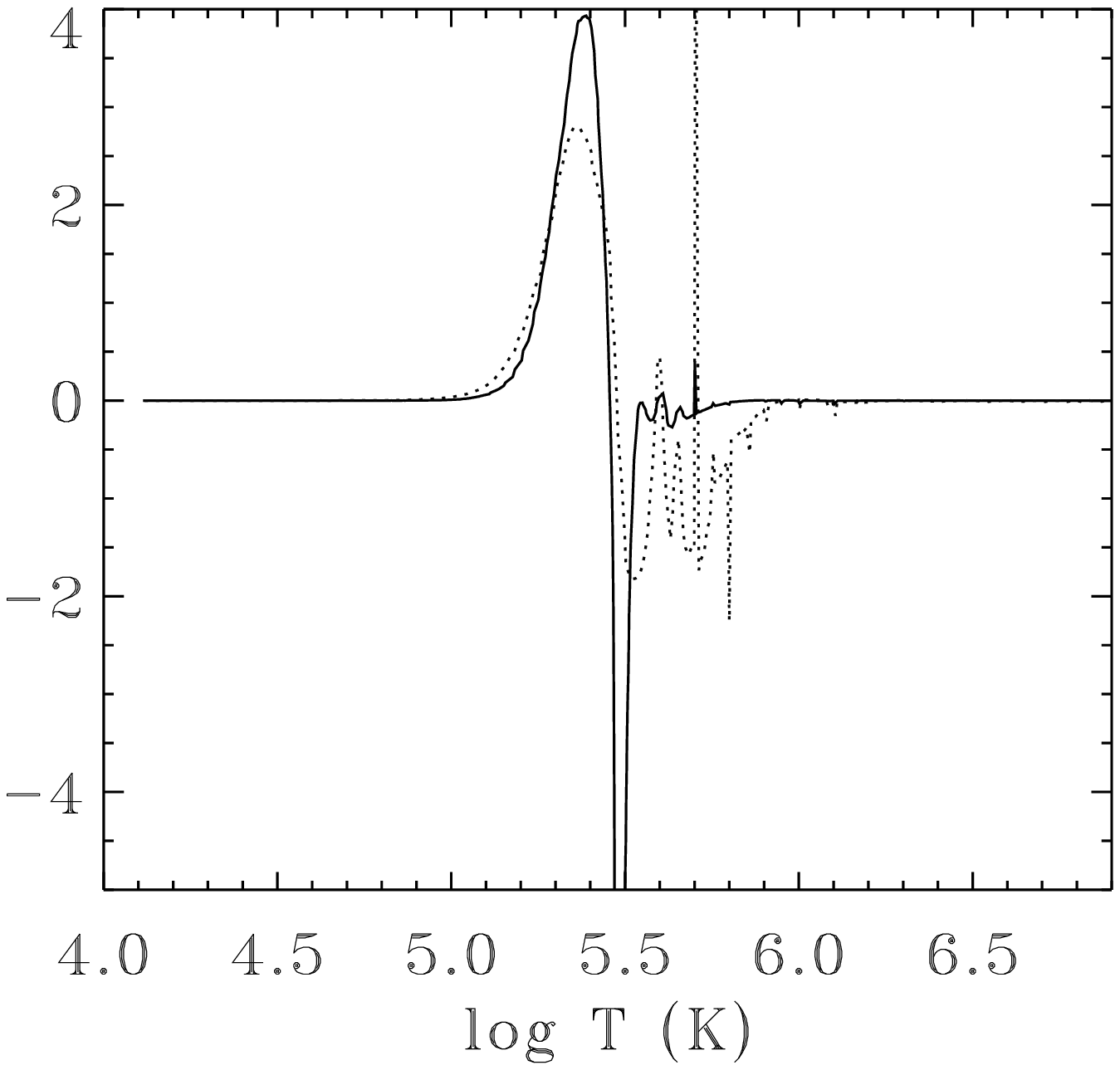} \\
 a. opacity & b. integrand of the work integral
 \end{tabular}}
\caption[]{The left-hand side shows the logarithm of the Rosseland
opacity with respect to temperature. The solid line is for a 4~M$_\odot$
model at 100~Myr with diffusion while the dotted curve is a similar model
with practically no diffusion. The right-hand side shows the integrand of the work
integral for the same models for a $\ell=1, n=26$ gravity mode. Pressure
modes and low order radial modes are damped while high order gravity
modes are excited. In this case both the mode is overstable in both
models but with a slightly larger growth rate in the model with
diffusion.}
\label{fig:opac}
\end{figure}

The increase in opacity in the model with diffusion is illustrated 
on the left-hand side of Fig.~\ref{fig:opac} and the resulting changes
in the integrand to the work integral is shown on the right-hand side.
Though the opacity differs in both models throughout the upper
envelope, the work integral for this and other high order g-modes 
depends very little on the regions cooler $\log T<5.0$. This may imply
that these modes may not be damped efficiently at low temperature that
will be discussed in the next section.

Adding mixing to prevent the formation of abundance stratification
in the models, would ensure that solar metallicities would 
be retained and one would therefore expect driving as it occurs in SPB stars
leading to a normal fraction of variable HgMn stars, {\sl i.e.} roughly 50\% 
of HgMn stars in the instability
region would be variable with amplitudes of approximately 10 mmag.

\section{Reconciling the Models with the Observations}

As the models are contradicted by observations in B stars
it is obvious that some  additional physical process or processes need to be included in the
models either to reduce the metallicity in the driving region or to
increase damping elsewhere.

It is fair to ask if the models with diffusion are representative of HgMn stars.
In fact they do not reproduce the surface anomalies of typical HgMn
stars very well. The models require that the entire upper envelope,
from the iron convection zone to the photosphere, be mixed to avoid
numerical problems. This is unlikely to the case in real stars. The very large and very varied anomalies
found in HgMn stars most likely show that diffusion occurs in the atmosphere. 
In fact, it this difficult to argue that surface abundances tell us much
about the internal composition of HgMn stars. It is therefore unlikely that the 
abundance profiles in our models reproduce well those in real HgMn stars.
As a result the opacity profile can also be expected to be different.

Models with varying homogeneous chemical abundances in the superficial
mixed zone have been examined to infer the effect of abundance
variations in possibly damping pulsations. No hints that it may be case
has been found but a more detailed analysis is needed to rule out this
possibility.

Mass loss could also be invoked to explain the differences between SPBs
and HgMn stars. Mass loss can remove surface abundance anomalies and
after some time could possibly empty the reservoir of iron-peak elements
in the driving region for pulsations. It would require a fine tuning of
the mass loss rate to remove chemical anomalies in SPB stars in a short
time scale (while remaining unobserved), and a metallic mass loss 
\cite{Babel96} fast enough in some elements to lower the opacity in the driving
region, and slow enough for other elements so they can accumulate in the
atmosphere.


\acknowledgements
This work was performed in part under the auspices of the U.S.
Department of Energy, National Nuclear Security Administration by the
University of California, Lawrence Livermore National Laboratory under
contract No.W-7405-Eng-48. Olivier Richard thanks G. Michaud for his
financial support. We thank
the R\'eseau Qu\'ebecois de Calcul Haute Performance (RQCHP) for
providing us with the computational resources required to compute
the stellar models.

\end{article}

\begin{thebibliography}{}

\bibitem[\protect\citeauthoryear{Adelman}{1998}]{Adelman98}
Adelman, S. J.
\newblock {On the HIPPARCOS photometry of chemically peculiar B, A, and
F stars}.
\newblock {\em A\&A Suppl. Ser.}, 132:93-97, 1998.

\bibitem[\protect\citeauthoryear{Adelman et al.}{2002}]{Adelmanetal02}
Adelman, S. J., A. F. Gulliver, O. P. Kochukov, and T. A. Ryabchikova.
\newblock {The Variability of the HgII $\lambda$3984 Line of the
Mercury-Manganese Star $\alpha$ Andromed{\ae}}.
\newblock {\em ApJ}, 575:449-460, 2002.

\bibitem[\protect\citeauthoryear{Babel}{1996}]{Babel96}
Babel, J.
\newblock {The fading of radiatively driven winds in B stars}.
\newblock {\em A\&A}, 309:867-878, 1996.


\bibitem[\protect\citeauthoryear{Pamyatnykh}{1999}]{Pamyatnykh99}
Pamyatnykh, A. A.
\newblock {Pulsational Instability Domain in the Upper Main Sequence}.
\newblock {\em Acta Astronomica}, 49:119--148, 1999.

\bibitem[\protect\citeauthoryear{Richard et al.}{2001}]{Richardetal01}
Richard, O., G. Michaud, and J. Richer.
\newblock {Iron Convection Zones in B, A, and F Stars}.
\newblock {\em ApJ}, 558:377--391, 2001.

\bibitem[\protect\citeauthoryear{Richer et al.}{2000}]{Richeretal00}
Richer, J., G. Michaud, and S. Turcotte.
\newblock {The Evolution of AmFm Stars, Abundance Anomalies, and
Turbulent Transport}.
\newblock {\em ApJ}, 529:338--356.

\bibitem[\protect\citeauthoryear{Vauclair \& Vauclair}{1982}]{VauclairVauclair82}
Vauclair, S. and G. Vauclair.
\newblock {Element Segregation in Stellar Outer Layers}.
\newblock {\em Ann. Rev. Astron. and Astrophys.}, 20:37--60.

\end{thebibliography}
\end{document}